\documentclass[10pt]{article}
\usepackage[dvips]{graphics}
\usepackage{amsfonts}
\newcommand{\Rl}{\mathbb{R}}

\newcommand{\Ir}{\mathbb{Z}}
\newcommand{\Cx}{\mathbb{C}}
\newcommand{\HH}{\mathcal{H}}
\newtheorem{theorem}{Theorem}[section]          % Numbering by sections
\newcommand{\Section}[2][{ }]%
{\section[#2]{#2}\setcounter{equation}{0}\setcounter{theorem}{0}
\setcounter{figure}{0}}

\newcounter{saveeqn}

\newcommand{\eq}[1]{(\ref{#1})}

\newcommand{\ba}[1]{\begin{array}{#1}}
\newcommand{\ea}{\end{array}}

\newcommand{\be}{\begin{equation}}
\newcommand{\ee}{\end{equation}}
\newcommand{\bea}{\begin{eqnarray}}
\newcommand{\eea}{\end{eqnarray}}
\newcommand{\beann}{\begin{eqnarray*}}
\newcommand{\eeann}{\end{eqnarray*}}

\newcommand{\up}{\uparrow}

\newcommand{\ua}{\hspace{3pt}$\uparrow$\hspace{3pt}}
\newcommand{\da}{\hspace{3pt}$\downarrow$\hspace{3pt}}
\newcommand{\ket}[1]{\left\vert #1\right\rangle}

\def\idty{{\leavevmode{\rm 1\ifmmode\mkern -5.4mu\else
\kern -.3em\fi I}}} \def\Ibb #1{ {\rm I\ifmmode\mkern -3.6mu\else\kern
-.2em\fi#1}} \def\Ird{{\hbox{\kern2pt\vbox{\hrule height0pt depth.4pt
width5.7pt \hbox{\kern-1pt\sevensy\char"36\kern2pt\char"36} \vskip-.2pt
\hrule height.4pt depth0pt width6pt}}}}
\def\Irs{{\hbox{\kern2pt\vbox{\hrule height0pt depth.34pt width5pt
\hbox{\kern-1pt\fivesy\char"36\kern1.6pt\char"36} \vskip -.1pt \hrule
height .34 pt depth 0pt width 5.1 pt}}}}
\def\QED{{\hspace*{\fill}{\vrule height 1.8ex width 1.8ex }\quad} \vskip
0pt plus20pt} 
\def\phi{\varphi}            % redefinition!
\def\epsilon{\varepsilon}    % redefinition!

\begin{document}
{
% The next 5 lines are for archived version only:
\baselineskip=10pt \thispagestyle{empty}
\noindent{Archived as {\tt math-ph/9809027}
and {\tt mp\_arc yy-nnn} \hspace{\fill}\newline
Preprint CPT-MNX-98}. September 30, 1998.\par
% Version of \today.\par
\vspace{30pt}
\noindent
{\Large\bf Interface states of quantum spin systems\footnote{Copyright 
\copyright\ 1998 by the authors. Reproduction of this article by any 
means is permitted for non-commercial purposes.}
\normalsize\rm}
\vspace{24pt}\par
\noindent{\bf Alain MESSAGER$^1$ and Bruno NACHTERGAELE$^2$}
\vspace{12pt}\par
\it\noindent
$^{(1)}$ Centre de Physique Th\'eorique, CNRS-Luminy, Case 907,
F-13288 Marseille, France
\newline
\phantom{$^(1)$} E-mail: {\tt messager@cpt.univ-mrs.fr}\newline
$^{(2)}$ Department of Mathematics, University of California,
Davis, Davis, CA 95616-8633, USA\newline
\phantom{$^(2)$} E-mail: {\tt bxn@math.ucdavis.edu}
\rm

\vspace{24pt}\footnotesize
\begin{quote}
{\bf Abstract:} We review recent results as well as ongoing work
and open problems concerning interface states in quantum spin
systems at zero and finite temperature.\par
\vspace{3pt}
{\bf Keywords:} Heisenberg ferromagnet, XXZ model, ground state selection, 
rigidity of interfaces, 111 interface, quantum SOS model
\end{quote}
\normalsize
}

\Section{Introduction}

Interface states have been extensively studied in rigorous
statistical mechanics since Dobrushin's seminal paper on interfaces
Gibbs states in the three-dimensional Ising model \cite{Dob}.
Interfaces and surfaces are everywhere in the physical world, and
to know their properties is important in understanding phenomena as
diverse as droplet evaporation, crystal growth, catalysis, dynamics
of magnets, etc.. Until recently  almost all rigorous
work was restricted to classical models, while some of the
phenomena are certainly more adequately described by quantum
mechanical models. In the past few years, however, several
interesting results have been obtained for the quantum Heisenberg
and the Falicov-Kimball model.  The Falicov-Kimball model is
discussed in several other contributions in the this volume
(\cite{Dat_h,Gru_h,Ken_h}.) Therefore, we restrict our attention
here to the Heisenberg model and quantum spin systems in general.
We will also introduce a quantum solid-on-solid (QSOS) model for
the 111 interface of the Heisenberg XXZ ferromagnet.

\Section{Interface ground states of the XXZ ferromagnets}

A quantum spin  model is defined by specifying a dynamics on an algebra of
quasi-local observables. The local structure is given by the finite
subsets of the $d-$dimensional latttice $\Ir^d$. With each site $x\in
\Ir^d$, there is associated a copy $\mathcal{A}_x$ of the $n\times n$
matrices with complex entries $M_n(\Cx)$. $\mathcal{A}_x$ is the algebra
of observables
at the
site $x$. For every finite subset $\Lambda\subset\Ir^d$, the observables
in the volume $\Lambda$ are given by
$$
\mathcal{A}_\Lambda=\bigotimes_{x\in\Lambda}\mathcal{A}_x
$$
This is a finite-dimensional $C^*$-algebra, and if
$\Lambda_0\subset\Lambda$, we have the natural embedding
$$
\mathcal{A}_{\Lambda_0}=\mathcal{A}_{\Lambda_0}
\otimes\idty_{\mathcal{A}_{\Lambda\setminus\Lambda_0}}
\subset\mathcal{A}_\Lambda
$$
The algebra of {\em local observables} is then defined by
$$
\mathcal{A}_{\rm loc}=\bigcup_{\Lambda\subset\Ir^d}\mathcal{A}_\Lambda
$$
where the union is over all finite subsets of $\Ir^d$.
Its completion is the $C^*-$algebra of {\em quasi-local observables}:
$$
\mathcal{A}=\overline{\mathcal{A}_{\rm loc}}
$$

We will also need the translation automorphsims on $\mathcal{A}$, denoted
by $\tau_x,x\in\Ir^d$, canonically mapping $\mathcal{A}_\Lambda$ into
$\mathcal{A}_{\Lambda +x}$.

The dynamics is determined by a family of {\em local Hamiltonians}.
For simplicity we will only discuss models with translation invariant
finite range interactions. I.e., let $h=h^*\in\mathcal{A}_{\Lambda_0}$,
for some finite set $\Lambda_0$, and define the local Hamiltonians by
$$
H_\Lambda=\sum_{x:\Lambda_0+x\subset\Lambda}\tau_x(h)
$$
The generator of the dynamics is the unique closed extension of the
derivation
$$
\delta(A):=\lim_{\Lambda\uparrow\Ir^d}[H_\Lambda,A],
\quad\mbox{ for all } A\in\mathcal{A}_{\rm loc}
$$
This generator can be exponentiated to obtain a strongly continuous
one-parameter group of $C^*$-automorphisms $\{\alpha_t\}_{t\in\Rl}$,
$$
\alpha_t(A):=e^{it\delta}(A),\quad A\in\mathcal{A}
$$
The standard proof can be found in the books by Bratteli and Robinson
\cite{BR} or Simon \cite{Sim}.

A {\em state} of the quantum spin system is a linear functional
$\omega$ on $\mathcal{A}$ with the properties:
$$
\omega(A^*A)\geq 0,\quad\mbox{ for all } 
A\in \mathcal{A},\quad \omega(\idty)=1
$$
The {\em ground states} of a model, by definition,
are the solutions of the following set of inequalities:
$$
\omega(A^*\delta(A))=\lim_{\Lambda\uparrow\infty}\omega(A^*[H_\Lambda,A])
\geq 0, \quad\mbox{ for all } A\in \mathcal{A}_{\rm loc}
$$
These inequalities express that all local excitations, i.e., created by
a local observable $A$, raise the energy.
This is also called {\em Local Stability} and we will often refer
to this set of inequalities as LS.

For translation invariant states, i.e., states satisfying
$$
\omega\circ\tau_x=\omega, \quad\mbox{ for all }x\in\Ir^d
$$
the ground states are exactly the states that minimize the energy per site:
$$
\omega(h)=\inf\{\eta(h)\mid \eta \mbox{ translation invariant state on
}\mathcal{A}\}
$$
In general, LS has non-translation invariant solutions, e.g., describing
domain walls or interfaces. In this paper, our main interest are these
non-translation invariant solutions. In one dimension they are often
called kink (and antikink) states, or soliton states. Solutions of LS
are commonly constructed as limits as $\Lambda\to\Ir^d$, of ground states of
the finite-volume Hamiltonians plus boundary terms.

One-dimensional quantum spin models, also called spin chains, are of
particular interest for several reasons. First of all it is in one
dimension that the most detailed rigorous analysis is possible. This was
the main motivation for \cite{LMat}. More recently it has also become
possible to realize quantum spin chains experimentally and to compare theory
and experiment in surprising detail. It is also
interesting that there is a special but not so small class of
one-dimensional models for which the exact ground states can be given
explicitly \cite{AKLT,FNW1,FNW2,Nac1}. And finally there are the integrable
quantum spin chains and the rich mathematical structures they exhibit
\cite{JM}.

None of the works cited above, however, deal with non-periodic ground
states. Consequently, the problem of characterizing the complete set
of solutions of LS, for any model, was, until recently, never dealt with.

The most detailed results have been obtained for the ferromagnetic
XXZ Heisenberg model. For $x\in \Ir^d$, let $S^i_x$, $i=1,2,3$, be
the standard spin-$S$ matrices, generating a $n=2S+1$-dimensional
irreducible unitary representation of SU(2), with
$S=1/2,1,3/2,\ldots$. The local Hamiltonians of the ferromagnetic
XXZ Heisenberg model are given by
\be
H_\Lambda=-\sum_{x,y\in\Lambda\atop \vert x-y\vert =1}
\frac{1}{\Delta}(S_x^{1} S_y^{1} + S_x^{2} S_y^{2})
+ S_x^{3} S_y^{3}
\label{hamxxz}\ee
with $\Delta \geq 1$. $\Delta = +\infty$ is the Ising model.
$\Delta=1$ is the isotropic model, also called the XXX model.

First, we discuss the complete set of ground states, in
the sense of LS for the XXZ ferromagnetic chain, i.e., $d=1$.
This class of models depends on two parameters: the dimension of the spin
matrices ($2S+1, S=1/2,1,3/2,\ldots $), and the anisotropy $\Delta\geq 1$.

Consider finite volumes of the form $\Lambda=[a,b]\subset\Ir$, and
denote by $\partial\Lambda=[a,a+r]\cup[b-r,b]$, the boundary
of $\Lambda$, where $r\geq 0$ is a suitably chosen integer.

Solutions of the ground state inequalities can be
constructed by adding suitable boundary terms $b_\Lambda
\in\mathcal{A}_{\partial\Lambda}$ to the
finite-volume Hamiltonians, and taking limits $\Lambda\uparrow\Ir$
of finite volume ground states of the form
$$
\omega_\Lambda(A)=\frac{\langle\psi_\Lambda, A\psi_\Lambda\rangle}{
\langle\psi_\Lambda,\psi_\Lambda\rangle}
\,\quad\mbox{ for all } A\in \mathcal{A}_\Lambda
$$
where $\psi_\Lambda\in\bigotimes_{x\in\Lambda}\Cx^{2S+1}$,
is an eigenvector belonging to the smallest eigenvalue
of $H_\Lambda +b_\Lambda$.

For the XXZ chains it suffices to take $r=1$ and boundary terms of the
form
\be
b_{[a,b]}=B(S^3_{b+1}-S^3_{a-1})
\label{bc}\ee
For convenience we will consider the chain on intervals of the form
$[-L,L]$, including the boundary points ($-a=b=L-1$).

Two translation invariant solutions are trivial to find and have been
well-known for a long time: the unique states $\omega_\uparrow$ and
$\omega_\downarrow$ determined by
\beann
&&\omega_\uparrow(S^3_x)=\phantom{-}S,\quad \mbox{ for all } x\in \Ir\\
&&\omega_\downarrow(S^3_x)=-S,\quad \mbox{ for all } x\in \Ir
\eeann

By taking $B:=B(\Delta)\pm S\sqrt{1-1/\Delta^2}$, Alcaraz, Salinas, and 
Wreszinski \cite{ASW}, and independently, Gottstein and Werner \cite{GW},
found non-translation invariant solutions. A first parameterization of these
kink states is by the third component of the total spin, i.e., the eigenvalue
$M$ of $\sum_{x=-L}^L S^3_x$, which commutes with the Hamiltonian. The
possible values of $M$ are $-S(2L+1),-S(2L+1)+1,\ldots,S(2L+1)$. For each
value of $M$ there is exactly one ground state $\psi_M$. It is useful to
define a generating function, $\phi(z),z\in\Cx$,
\be
\phi(z)=\sum_M a_m z^M \psi_M
\label{generating_phi}\ee
where the $a_M$ are suitable constants. We refer to \cite{ASW,GW} for more
details. The important fact is that the $a_M$ can be chosen such that
$\phi(z)$ becomes a product:
\be
\phi(z)=\bigotimes_{x=-L}^L \chi(zq^{-x})
\label{generating_chi}\ee
with
$$
\Cx^{2S+1}\ni \quad\chi(z)=\sum_{m=-S}^S z^m q^{x(S-m)}
\sqrt{\frac{(2S)!}{(S-m)!(S+m)!}}\ket{m}
$$
where $S^3\ket{m}=m\ket{m}$, and $q\in (0,1)$, such that
$\Delta=(q+q^{-1})/2$. Note that, in the limit $L\to\infty$, a translation
by one lattice unit transforms $\phi(z)$ into $\phi(zq)$. Therefore, we can
use $\phi(z)$ as a parameterization of the kink states by translations.
It is then a trivial computation to get the magnetization profile in
the states $\phi(z)$. E.g., for $S=1/2$, one finds 
\beann
\omega_{\rm kink}(S^3_x)&=&\phantom{-}\frac{1}{2}\tanh ((x-a)/\xi)\\
\omega_{\rm antikink}(S^3_x)&=&-\frac{1}{2}\tanh ((x-a)/\xi)
\eeann
where $\xi$ and $a$ are simple functions of $q$ and $z$.

The concept of {\em zero-energy states} plays an important role in
the characterization of the complete set of solution of LS. So, we
introduce it here, before we continue our discussion of the XXZ model.

Suppose there exists $0\leq\tilde{h}\in\mathcal{A}_{\Lambda_0}$, such that
$$
H_\Lambda + b_\Lambda :=\tilde{H}_\Lambda
=\sum_{x:\Lambda_0+x\subset\Lambda}\tau_x(\tilde{h})
$$
Then, if a state $\omega$ satisfies
$$
\omega(\tau_x(\tilde{h}))=0,\quad \mbox{ for all } x\in\Ir,
$$
$\omega$ is called a {\em zero-energy state}. It is easy to
show that any zero-energy state satisfied LS.

Gottstein and Werner obtained the complete set of zero-energy states of the
anisotropic XXZ chain $\Delta >1$ with the particular boundary terms
given in \eq{bc}. They proved that any pure zero-energy state is either one
of the two translation invariant ones, or it is a member of a set of
mutually equivalent kink states or a set of mutually equivalent antikink
states. A state $\omega$ is called {\em pure}, iff for any two states
$\omega_1,\omega_2$, and $t\in (0,1)$, one has
$$
\omega=t\omega_1+(1-t)\omega_2\Rightarrow \omega_1=\omega_2
$$
It is obvious that the solution set of LS is convex, and one can prove it
is a face.  The same holds for the set of zero-energy states of a fixed
$\tilde{h}$. Therefore, finding the pure solutions is enough.

To which class $\omega$ belongs is determined by the limits
$\alpha,\beta\in\{\pm S\}$, i.e., its asymptotic behavior:
$$
\alpha:=\lim_{x\to-\infty}\omega(S^3_x),\quad
\beta:=\lim_{x\to+\infty}\omega(S^3_x)
$$
The following table summarizes the four types of zero-energy
ground states as parametrized by $\alpha$ and $\beta$ ($\Delta >1$):

\begin{center}
\begin{tabular}{|c||l|l|}\hline
$\alpha\;\beta$ & type &dominating configuration\\ \hline\hline
$++$&up&$\vert\cdots$\ua\ua\ua\ua\ua\ua\ua\ua\ua\ua\ua\ua$\cdots\rangle$\\
\hline
$--$&down&$\vert\cdots$\da\da\da\da\da\da\da\da\da\da\da\da$\cdots
\rangle$\\ \hline
$-+$&kink&$\vert\cdots$\da\da\da\da\da
$\cdots$\ua\ua\ua\ua\ua\ua$\cdots\rangle$\\ \hline
$+-$&antikink&$\vert\cdots$\ua\ua\ua\ua\ua $\cdots$
\da\da\da\da\da\da$\cdots\rangle$\\ \hline
\end{tabular}
\end{center}

We refer to \cite{ASW,GW} for more details on the kink and
antikink states. The case $\Delta=1$ will be discussed further on.

Let us now return to the question of characterizing {\em all\/} solutions
of Local Stability. The first result is due to Matsui \cite{Mat2}.

\begin{theorem}[Matsui, 1996]
For the ferromagnetic XXZ chain, with $\Delta >1$, and $S=1/2$,
the translation invariant states $\omega_{\uparrow}$ and
$\omega_{\downarrow}$,
together with the kink and antikink states described in \cite{GW}
and discussed above, are the full set of pure solutions of LS.
\end{theorem}

Note that this theorem does not say anything about the isotropic
model ($\Delta =1$). The isotropic ferromagnetic chains have an
infinite family of translation invariant ground states, given
by the state $\omega_\up$ and all rotations of it. As expected,
this breaking of the rotation symmetry is accompanied by a gapless
excitation spectrum. The proof of the above theorem relies on
the existence of gap, and, therefore, does not extend to the isotropic
case. Recently, Koma and Nachtergaele obtained a proof of the complete set
of ground state for all values of $S$ and $\Delta \geq 1$
\cite{KN2}.

\begin{theorem}[Koma and Nachtergaele, 1997]
For the ferromagnetic XXZ chain, with $\Delta >1$, and $S\geq 1/2$,
the translation invariant states $\omega_{\uparrow}$ and
$\omega_{\downarrow}$,
together with the kink and antikink states generalized to arbitrary $S$,
as in \cite{ASW} exhaust the set of pure solutions of
LS. If $\Delta=1$, all solutions are translation invariant.
\end{theorem}

So, in the isotropic case there are no kink-type ground states.
One expects that the same is true for models with a unique translation
invariant (or periodic) ground state. This has not yet been proved
in general, but Matsui \cite{Mat3} obtained a quite general
result in the case of a unique zero-energy state, which we explain
next.

Consider an arbitrary spin chain with a translation invariant
nearest neighbor interaction, i.e., $h\in\mathcal{A}_{[0,1]}$ :
$$
H_{[a,b]}=\sum_{x=a}^{b-1} \tau(h)
$$
Assume that $h\geq 0$ and that there is a unique translation invariant
state $\omega$ such that
$$
\omega(h_{x,x+1})=0,\quad\mbox{ for all } x\in\Ir
$$
One would expect that in this case $\omega$ is the unique solution of LS.
Matsui's result, the only result so far, requires an additional assumption on
the set of zero-energy states of the the half-infinite chains, i.e., states
$\eta$ on $\mathcal{A}_{[1,+\infty)}$ ($\mathcal{A}_{(-\infty,0]}$,
respectively) such that
$$
\eta(h_{x,x+1})=0,\quad\mbox{ for all } x>0\quad  (x<0,\ \mbox{ resp.}).
$$

\begin{theorem}[Matsui, 1997]
For a spin chain as described above, if $\omega$ is the unique zero-energy
ground state and if all zero-energy states of the left and right
half-infinite chains are quasi-equivalent, then $\omega$ is the unique
solution of LS.
\end{theorem}

In particular this implies that the AKLT spin 1 chain introduced
in \cite{AKLT} has a unique ground state in the sense of
LS. 

In higher dimensions there also exist non-translation invariant ground
states. In finite volume, this was already noted  in \cite{ASW}.
Koma and Nachtergaele proved that ground states with a rigid interface
in the diagonal (i.e., $11\cdots 1$) direction exist in all dimensions.
They also proved that the excitation spectrum above the ground
state with a 11 interface in two dimensions is gapless
\cite{KN4}. Matsui generalized this to arbitrary dimensions $d\geq 2$
\cite{Mat4}. We now give an easy construction of a 111 interface state
for the XXZ ferromagnet.

\begin{figure}[t]
\begin{center}
\resizebox{!}{7truecm}{\includegraphics{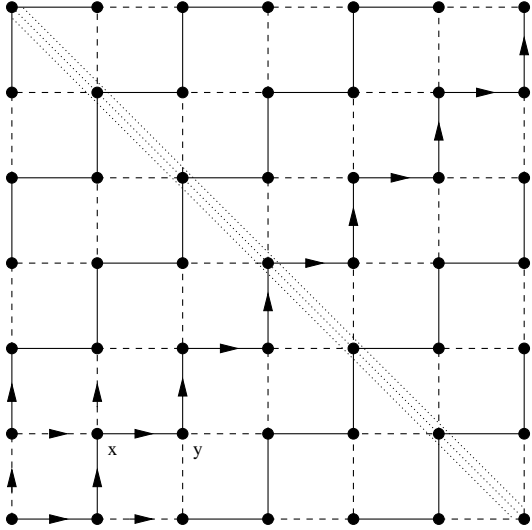}}
\parbox{14truecm}{\caption{\baselineskip=14 pt\label{fig:diagint}
The decomposition of the square lattice into coupled zig-zag chains.
The full lines are the bonds in the chains. The dashed lines are bonds
carrying the interchain couplings. The triple dotted line indicates the
position of the diagonal interface. All vertical bonds are oriented upwards,
and all horizontal bonds are oriented to the right.}}
\end{center}
\end{figure}

It follows from \eq{generating_chi} that, for all $z\in\Cx$,
$$
h\; \chi(z)\otimes\chi(zq^{-1})=0,
$$
where
\be
h:=-(S^1\otimes S^1+S^2\otimes S^2)-\Delta(S^2\otimes S^3 -S^2)
+B(\Delta)(S^3\otimes\idty-\idty\otimes S^3)
\label{hxy}\ee
is the positive definite interaction for which the kink states
are the zero-energy ground states.
Let $h_{x\to y}$ denote a copy of $h$ acting on the Hilbert space
$\HH_{\{x,y\}}$ of the sites $x$ and $y$ {\em in this order\/}.
It follows immediately that, for any $\Lambda\subset\Ir^d$,
$x,y\in \Lambda$, such that $\vert x-y\vert =1$, we have
$$ 
h_{x\to y} \Omega(z)=0
$$
with
$$
\Omega(z):=\bigotimes_{u\in\Lambda} \chi(zq^{-\vert u\vert})
$$
Hence,
\be
H_\Lambda\Omega(z)
=\sum_{x,y\in\Lambda, \vert x\vert<\vert y\vert\atop\vert x-y\vert=1}
h_{x\to y} \Omega(z)=0
\label{ham_d}\ee
With the observation that every interior point in $\Lambda$ is the starting
point ($x$) and the end point ($y$) of an equal number of bonds,  it is easy
to see that $H_\Lambda$ is the XXZ Hamiltonian with a boundary field. We
conclude that $\Omega(z)$ is a zero-energy ground state for the XXZ model on
$\Ir^d$ for any  $d$.

For concreteness, let us continue the discussion in the two-dimensional
case. Generalization to arbitrary dimensions is straightforward.
The two-dimensional system can be decomposed into a collection of
coupled one-dimensional XXZ models defined the zig-zag chains shown
as full lines in Figure \ref{fig:diagint}. The bonds coupling two
zig-zag chains together are indicated by dashed lines.
As $\Omega(z)$ is a zero-energy ground state of the two-dimensional
system, its restriction to a zig-zag line must be a zero-energy ground
state of the one-dimensional XXZ model. Note that the boundary terms
coincide with the ones used before to obtain the kink ground states
if we interpret the orientations of the bonds in two dimensions as
left-to-right in one dimension.
Further inspection shows that the kink states on the zig-zag
lines are aligned such that they are all centered on a line in the main
diagonal direction (11) of the two-dimensional square lattice.
Thus, $\Omega(z)$ describes a $11\cdots$1-interface.

\begin{theorem}[Koma and Nachtergaele, unpublished; Matsui 1997]
If $d\geq 2$, $S\geq 1/2$, and $\Delta>1$, the family of product states
$\Omega(z),z\in\Cx$, normalized and extended to infinite volume,
are all the pure zero-energy ground states of the interaction
defined in \eq{hxy}. The excitation spectrum above these
ground states is gapless.
\end{theorem}

We already noticed that $\vert z\vert$ determines the position of the kinks,
i.e., a change in $\vert z\vert$ corresponds to a translation of the
interface. It is not hard to see that the phase of $z$ can be changed
by applying another symmetry of the Hamiltonian:
$$
e^{i\theta S^3_{\rm tot}}\Omega(z)=\Omega(e^{i\theta}z)
$$
Rotations about the $z-$axis are commute with the Hamiltonian,
but the states $\Omega(z)$ are not invariant under it. Therefore,
we have identified a broken continuous symmetry that goes along with
the breaking of translation invariance in interface states.
This is the basis of Matsui's general proof of the existence
of gapless excitations above these interface states of the
XXZ model, along the lines of \cite{LPW}.

For sufficiently large $\Delta$, and $d\geq 3$, interface ground states
(as well as low-temperature states) with the interface along the main
coordinate directions (100), have also been shown to exist \cite{BCF1,BCF2}.
In this case, however, there is a gap above the ground state.
Another difference is that interfaces in the 100 directions can be
understood as perturbations of the same interface states for the Ising
model, which describes the limit $\Delta \to\infty$. 
The Ising model does not have a rigid 111 interface. It is easy to see
that the ground states with 111 boundary conditions is infinitely
degenerate. See \cite{DMN} for a discussion.
For finite $\Delta$, however, the $XY$ term in 
the XXZ Hamiltonian produces a ground state selection effect, i.e., 
lifts the degeneracy.

\Section{A quantum SOS model}

The next natural question concerns the existence of a rigid 111 interface
at non-zero temperature. The structure of a 111 interface in a lattice
model on $\Ir^3$ is already quite complicated. In the case of the XXZ
model, the presence of gapless excitations in the interface ground states,
is additional difficulty. In the past, interesting progress with interface
problems has been made by introducing models where the interface is
described by a height function. This eliminates most of the complicated
geometry of the interfaces, which usually does not substantially affect the
most relevant physical properties. Such models are called  Solid-On-Solid
(SOS) models. In this section, we introduce an quantum SOS (QSOS) model for
the 111 interface of the anisotropic Heisenberg ferromagnet. As this model
is a quantum system, the most straightforward and reliable way of obtaining
an approximation is by restricting the Hilbert space to a suitable
subspace. We would also like to retain as much as possible particular
features, as, e.g., the gapless excitations of the interface.

The restricted Hilbert space for the QSOS model should at least
contain the ground state interface, and capture the most important
excitations. In analogy with the classical SOS models we would
also like to have the property that each one-dimensional subsystem
defined on a line in the 111 direction, i.e., perpendicular to the
interface, is found in one of its ground states. In the classical
case this would mean that any such line crosses the interface
exactly once in any SOS configuration. This implies that the interface
can be described by a height function. In our case the one-dimensional
subsystems are naturally defined on the zig-zag lines drawn in full
in Figure \ref{fig:diagint}. This idea can easily be generalized to
3 and more dimensions. For each one-dimensional the natural Hilbert space
is then given by
\be
\HH_{Hz}=\mbox{closed linear span of }\{\phi(z)\mid z\in\Cx\}
\ee
By going back to the basis of kink states with fixed magnetization $M$,
one can easily show that this Hilbert space is also given by
$$
\HH_{\rm zz}=\mbox{closed linear span of }\{\phi(q^n)\mid n\in\Ir\}
\cong l^2(\Ir)
$$
The states $\phi(q^n)$, $n\in\Ir$, are an non-orthogonal basis for
$\HH_{\rm zz}$, and the index $n$ can be interpreted as the height function,
i.e., the position of the interface. There is a unitary operator $S$ on
$\HH_{\rm zz}$ such that
\be
S\phi(q^n)=\phi(q^{(n+1)})
\label{defS}\ee
The Hilbert space for the QSOS model is
then a tensor product of copies of $\HH_{\rm zz}$, one for each
zig-zag line:
$$
\HH_{QSOS}=\bigotimes_{u\in\Gamma} \HH_{{\rm zz},u}
$$
where $u\in\Gamma$ label the zig-zag lines. In the case $d=2$,
$\Gamma$ is a simple one-dimensional lattice. For $d=3$, $\Gamma$ is a
two-dimensional triangular lattice.

The next step in the definition of the QSOS model is the calculation of
the Hamiltonian. Let $P_{QSOS}$ denote the orthogonal projection
onto $\HH_{QSOS}$ considered as a subspace of the original Hilbert
space of the $d-$dimensional XXZ model. Then, the Hamiltonian of the
QSOS model is given by
$$
H_{QSOS}=P_{QSOS} H_{XXZ} P_{QSOS}
$$
Some symmetry properties follow immediately from this definition.
The first is invariance under a global shift of the interface. This is
expressed by the fact that
$$
[H_{QSOS},\bigotimes_{u\in\Gamma}S]=0
$$
where $S$ is the unitary operator defined by \eq{defS}. Because of this
invariance, one needs to consider suitable boundary conditions when taking
the thermodynamic limit in the directions parallel to the interface. One can
also implement this by defining the infinite-volume model on incomplete
tensor product in the sense of Guichardet \cite{Gui}. Furthermore, from the
definition of $\HH_{\rm zz}$ it is clear that $\HH_{QSOS}$ is invariant under
global rotations about the $z-$axis. Therefore, the rotation invariance of
$H_{XXZ}$ is inherited by $H_{QSOS}$. Finally, the translation invariance of
the original Hamiltonian,  which is also at the origin of the invariance
under $S$ (translations  perpendicular to the interface), leads to
periodicity or translation  invariance of the QSOS model over $\Gamma$.

There are different possibilities for the concrete representation
of $H_{QSOS}$. In order to represent this Hamiltonian as a matrix in an
orthonormal basis one needs to correct for the non-orthogonality of
the natural basis $\{\phi(q^n)\mid n\in\Ir\}$ of the $\HH_{\rm zz}$.
Although there are several examples in the physics literature where in the
definiton of projected models non-orthogonality of a natural set of basis
vectors is ignored, we think it is important to take it into account here.
We will elaborate on this in a future publication.
We conjecture that the QSOS model, with $d\geq 3$, has a rigid phase
at sufficiently low temperatures. We will report on partial results
on this problem in a future publication \cite{MN}.

\section*{Acknowledgments}
B.N. gratefully acknowledges hospitality at the CPT, where
part of the work reported on in this paper was carried out.
B.N. is partially supported by the National Science Foundation
(USA) under grant \# DMS-9706599.

\end{document}